\documentclass[twocolumn, a4paper, prl]{revtex4}

\usepackage{graphicx, epsfig, fancybox}
\usepackage{amssymb,amsmath}
\usepackage{color}
\usepackage{pdfsync}

\usepackage{times}

\def \del{\partial}    

\begin{document}




\title{Strongly interacting one-dimensional bosons in optical lattices of
arbitrary depth: From the Bose-Hubbard to the sine-Gordon regime and
beyond}

\author{Achilleas Lazarides}
\author{Masudul Haque}

\affiliation{Max Planck Institute for the Physics of Complex Systems,
N\"othnitzer Str.~38, 01187 Dresden, Germany}

\begin{abstract}

We analyze interacting one-dimensional bosons in the continuum, subject to a
periodic sinusoidal potential of arbitrary depth.  Variation of the lattice
depth tunes the system from the Bose-Hubbard limit for deep lattices, through
the sine-Gordon regime of weak lattices, to the complete absence of a lattice.
Using the Bose-Fermi mapping between strongly interacting bosons and weakly
interacting fermions, we derive the phase diagram in the parameter space of
lattice depth and chemical potential.  This extends previous knowledge from
tight-binding (Bose-Hubbard) studies in a new direction which is important
because the lattice depth is a readily adjustable experimental parameter.  
Several other results (equations of state, energy gaps, profiles in harmonic
trap) are presented as corollaries to the physics contained in this phase
diagram.  Generically, both incompressible (gapped) and compressible phases
coexist in a trap; this has implications for experimental measurements.

\end{abstract}


\pacs{67.85.-d, 05.70.Ln, 67.85.De, 67.85.Jk, 03.75.Kk, 03.75.Nt}


\maketitle

\emph{Introduction} ---
Ultracold atomic gases provide us with unprecedented experimental
opportunities for creating interacting quantum many-particle systems \cite{bloch,lewtool},
both in \emph{lattice} and in \emph{continuum} situations.  In response,
theoretical lattice and continuum models have been widely 
analyzed in the past decade and half.  However, it is experimentally
straightforward to \emph{interpolate} between the two cases, through periodic
optical potentials whose amplitude can be made to vary from zero to a large
value \cite{haller466pinning}.  The heating or cooling due to ramping up the
strength of the periodic potential is well-studied
\cite{heating_latticeloading}.  However, the potential well depth is generally
not considered as a parameter tuning the phase diagram, except indirectly via
effective Hubbard-model parameters which are strictly valid only in the
deep-potential limit.  The study of complex many-particle systems as
a function of the periodic potential depth, interpolating between continuum
and tight-binding limits, is a direction with many interesting effects whose
exploration is just beginning.  
The measurements of Ref.~\cite{haller466pinning} are promising first steps in
this direction.  This experiment focuses on one-dimensional (1D) physics,
where dimensional confinement enhances correlation and interaction effects.

\begin{figure}
\centering
\includegraphics*[width=0.99\columnwidth]{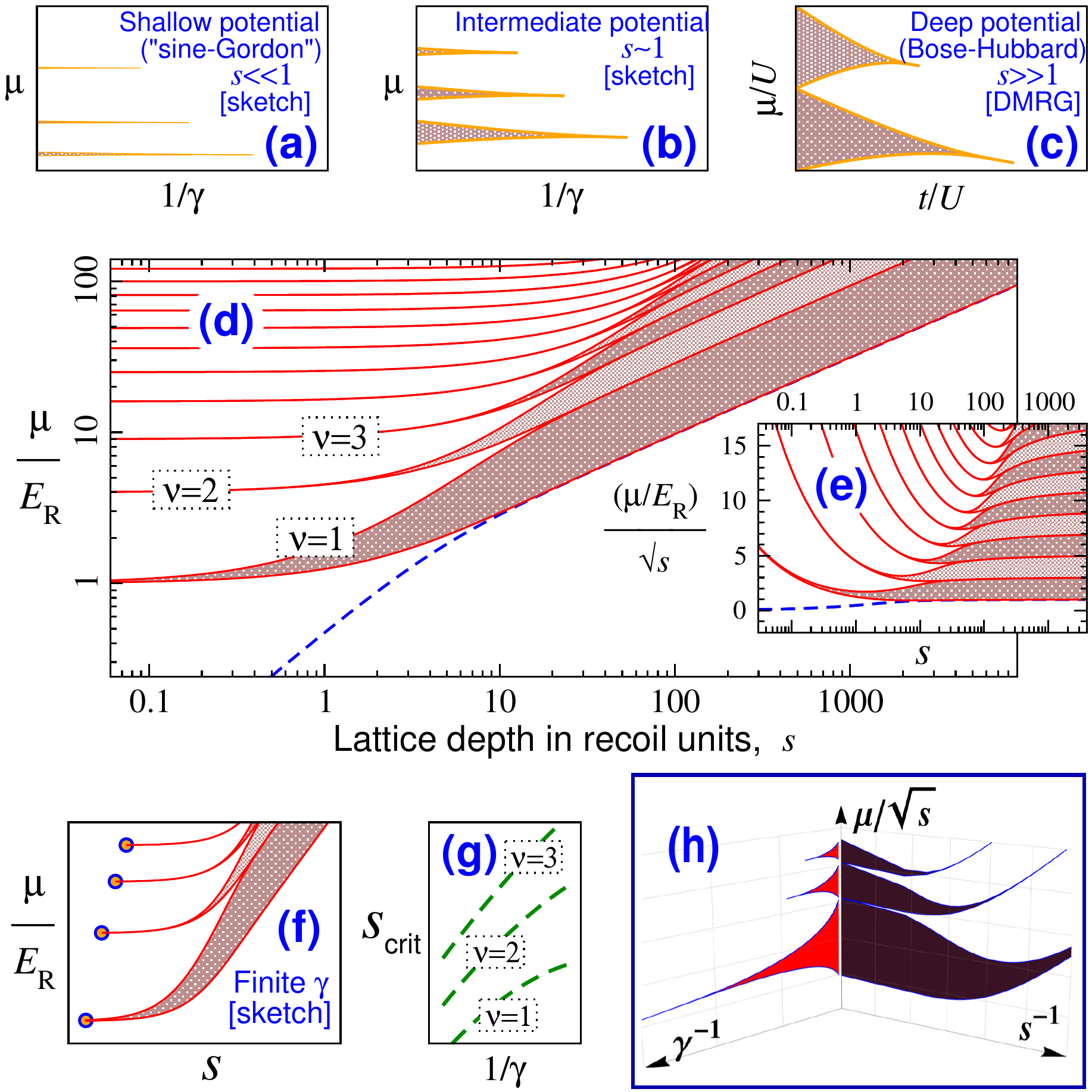}
\caption{ \label{fig_phasediagram} 
(a-c) Schematics of the $\mu$-$\gamma$ phase diagram for weak, intermediate
and deep lattices.  The Mott regions are shaded.  The deepest lattice is taken
to be described by a Bose-Hubbard model \cite{EjimaFehskeGebhard_EPL11}.
(d,e) Calculated $\mu$-$s$ phase diagram for the strongly interacting
($\gamma^{-1}=0$) limit.  Each Mott region corresponds to a different integer
filling $\nu$.  The dashed blue line is the energy of a single particle placed
in an empty system, i.e., the chemical potential for $\nu=0$.  (f,g) Inferred
schematics for finite $\gamma^{-1}$: phase diagram and critical depths.  (h)
Structure of the 3D phase diagram (in $\mu$-$\gamma$-$s$ space) showing the
`Bose-Hubbard' region (previously studied) and the large-$\gamma$  plane
(determined in this work).
}
\end{figure}

In this work, we address the system implemented in the experiment of
Ref.~\cite{haller466pinning}, namely, one-dimensional strongly interacting
bosons in the presence of a variable-strength optical potential.  We provide
the most fundamental information needed for understanding the physics in this
system, namely, the phase diagram.

Like many other systems, 1D bosons have been widely studied in the
tight-binding and continuum limits \cite{Cazalilla::2011}, but not much for
intermediate depths.  The deep-well limit is described by the Bose-Hubbard
model, which is a tight-binding model in the sense of each well being
described by a single mode.  The 1D Bose-Hubbard phase diagram is well-studied
theoretically \cite{Cazalilla::2011, batrouni1996world,freericks1994phase,
EjimaFehskeGebhard_EPL11}, and the model has been realized often
experimentally \cite{bloch,lewtool,Cazalilla::2011}.
In the other limit, i.e., without a periodic potential, the system is the
continuum Lieb-Liniger gas \cite{LiebLiniger_1960}, described by the
Hamiltonian $H=\sum_j\del^2/\del{x_j}^2 +g\sum_{i<j}\delta(x_i-x_j)$.  The
dimensionless interaction parameter is $\gamma=g m/(\hbar^2n)$, with $m$ the
mass of bosons and $n$ the one-dimensional density.  We are interested in
large $\gamma$, in or near the so-called Tonks-Girardeau (TG) regime.  The TG
gas has been intensively studied
theoretically \cite{Sen_pseudopotential,brand2005dynamic,Brand:PhysRevA72:2004,
kolomeisky, buljanlinear, cazalilla2004differences, Cazalilla::2011,
Cazalilla:PhysicalReviewA:2003,Dunjko:2001p10842,
Gangardt:NewJournalOfPhysics5:2003,Girardeau:1960p10745,paraan,
girardeauwright, Papenbrock_PRA03, MinguzziGangardt_PRL05}.  Some calculations
exist in the weak lattice limit, via mapping to a sine-Gordon field
theory~\cite{kehrein_PRL99, lazarides2011,Buchler:2003p10714}. The TG gas has
also been experimentally realized in a lattice version with low filling
\cite{expt_TG_lattice}.   Ref.~\cite{haller466pinning} studies strongly interacting 1D
bosons in the continuum, with the addition of a periodic potential of variable
strength.  Motivated by this, as well as by the dearth of theoretical studies
of the effect of the potential depth, we analyze a Tonks-Girardeau gas in a
periodic potential of variable depth.  The periodic potential is
$V(x)=sE_{R}\sin^{2}(qx)$, with lattice period $\pi/q$.  The parameter $s$
gives the potential depth in units of the recoil energy,
$E_{R}=\frac{\hbar^{2}q^{2}}{2m}$.

We use the Bose-Fermi transformation \cite{Girardeau:1960p10745} to map
strongly interacting 1D bosons ($\gamma\rightarrow\infty$) to free fermions.
The problem with a periodic potential then turns into a band-structure
problem, which we address using the Bloch theorem.  This rather simple setup
allows us to extract a remarkable amount of information.  The most prominent
result is the phase diagram in the chemical potential ($\mu$) versus well
depth ($s$) plane, i.e., we map out parameter regions which are Mott phases
and those which are superfluid phases [Fig.\ \ref{fig_phasediagram}(d,e)].
This phase diagram, whose structure to the best of our knowledge was not
previously known, is exact and detailed in the $\gamma\rightarrow\infty$
limit, and also provides a significant amount of information about the phase
diagram for finite but large $\gamma$.

We also present excitation gaps in the different Mott phases.  Defining a
``filling'' variable $\nu$ as the average density in a well, we present
equations of state, i.e, $\mu(\nu)$ curves, for various well depths.  In
addition, we calculate the effect of an arbitrary-strength periodic potential
in the presence of an overall harmonic trap.  This is important because, like
Ref.~\cite{haller466pinning}, foreseeable experiments are likely to be
performed in harmonic confinement.  Going beyond the infinitely interacting
case, we also provide results for large but finite interactions.

\emph{The phase diagram} ---
 Fig.\ \ref{fig_phasediagram}(a-c) illustrate the phase diagram in the
plane of chemical potential $\mu$ and (inverse) interaction strength, for
weak, intermediate and strong lattice strength $s$.  The deep-potential limit
has been represented as the well-studied Bose-Hubbard model (top right).  Away
from this limit, the phase diagrams are expected to be of the forms sketched,
but are not presently known in detail. 
%
%
It should be possible to determine parts of the
small-$s$ phase diagrams using the sine-Gordon model, but full results do not
yet exist in the literature.
%

Our calculation determines the small $\gamma^{-1}$ limit, i.e., the region
touching the vertical axes in each of the top-panel schematics, for all values
of the lattice depth $s$.  This can be put into context in terms ofthe 3D
phase diagram of Fig.\ \ref{fig_phasediagram}(h): our calculation maps out a
new plane in $\mu$-$\gamma$-$s$ space.

For $\gamma^{-1}=0$, the bosonic phase diagram is obtained from the
free-fermion band structure. A filled fermionic band corresponds to integer
filling in the bosonic system. Thus, the fermionic band-gap regions correspond
to Mott phase regions in our system.  The resulting exact phase diagram is
shown in Fig.\ \ref{fig_phasediagram}(d,e).  This phase diagram is our main
result.

The $s\rightarrow\infty$ region satisfactorily reproduces known features of
the Bose-Hubbard limit --- the Mott regions touch, killing superfluid regions,
and each occupies an equal range of $\mu$.  In the shallow ``sine-Gordon''
limit ($s\rightarrow{0}$), the Mott regions shrink to tiny slivers.  For
$\gamma^{-1}=0$ these Mott slivers continue down to $s=0$, so that an
infinitesimal periodic potential can pin a gas if the filling allows
commensurate pinning.  In the $s=0$ limit, the successive Mott slivers appear
at linearly increasing  gaps of $\mu$ from each other. 
%

In Fig.\ \ref{fig_phasediagram}(f) we sketch the expected phase diagram for
large but finite $\gamma$.  The Mott regions now start at some finite critical
lattice depth $s_{\rm crit}$.  The Mott regions are expected to be less robust
for larger fillings because of the smaller gaps and stronger perturbative
corrections at larger $\nu$ (see below).  It remains an open problem to
determine the $s_{\rm crit}(\gamma)$ curves for $\nu>1$, sketched
in \ref{fig_phasediagram}(g).  The $\nu=1$ curve can be obtained from a
sine-Gordon description \cite{haller466pinning}.  The other curves should be
be progressively higher [Fig.\  \ref{fig_phasediagram}(g)] but are otherwise unknown.

\begin{figure}
\centering
\includegraphics*[width=0.99\columnwidth]{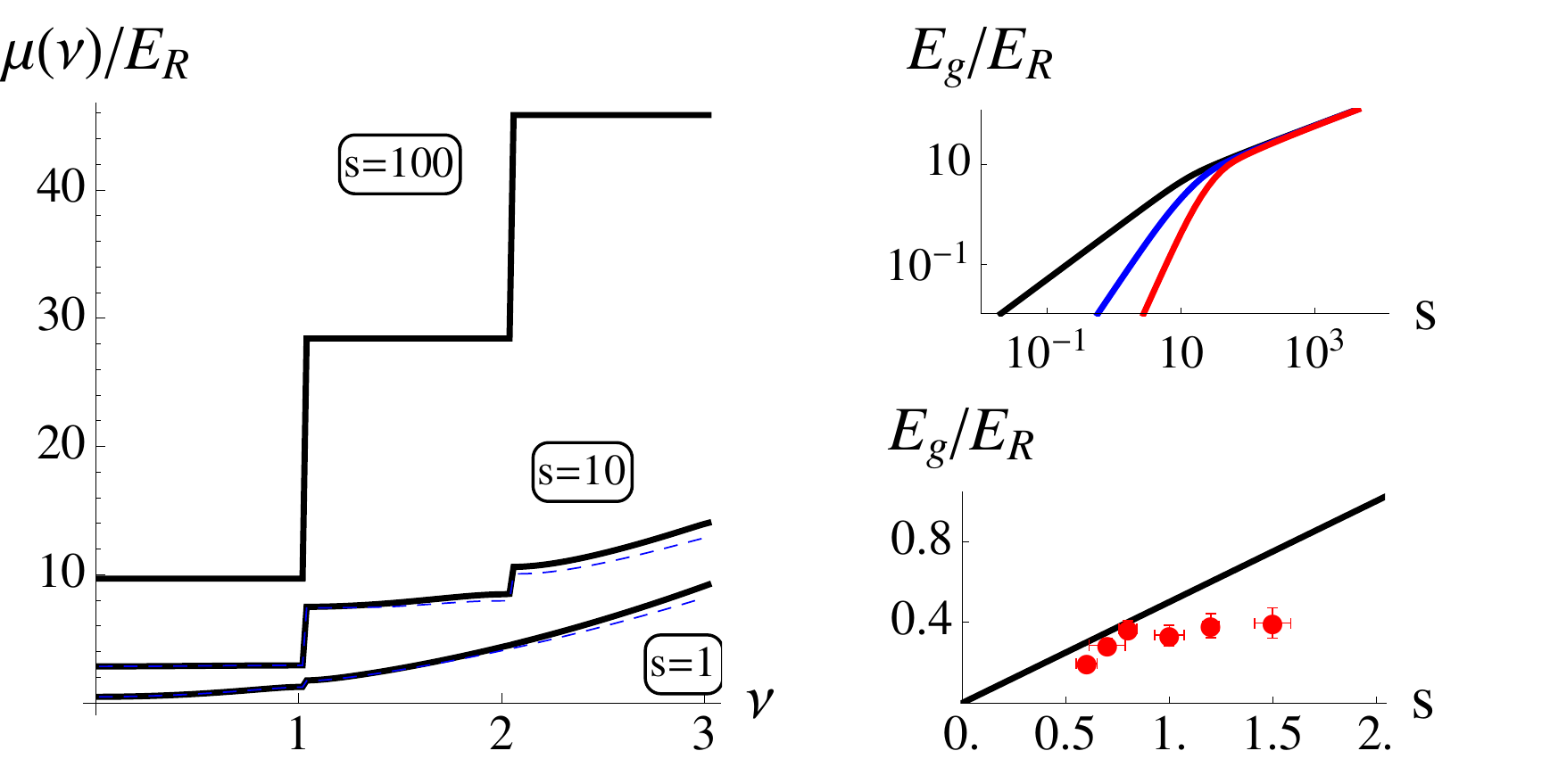}
\caption{ \label{fig_additional}
Left: Equation of state for various optical lattice depths. Solid curves
correspond to $\gamma\rightarrow\infty$, dashed curves to $\gamma=11$.  Right
top: energy gaps $E_{g}$ in $\nu=1$, $\nu=2$ and $\nu=3$ Mott phases (top
black, middle blue and bottom red) as a function of lattice depth.  The gaps
scale as $s^{\nu}$ at small $s$ and as $s^{1/2}$ at large $s$.  Right bottom:
calculated $\nu=1$ energy gap (line) shown with experimental values from
trapped-system measurements \cite{haller466pinning}.  }
\end{figure}

\emph{Bose-Fermi mapping} ---
To obtain the wavefunction of either $N$  hardcore bosons or $N$ noninteracting
fermions in an arbitrary single-particle 1D potential, we take the
first $N$ single-particle eigenfunctions $u_{i}(x)$. The many-body
wavefunction is then
$\psi_{f}(x_{1},x_{2},...,x_{N})=\sum_{P}\left(-1\right)^{P}\sum_{i}u_{i}(Px_{i})$
(where $P$ denotes a permutation) for fermions and
$\psi_{b}(x_{1},x_{2},...,x_{N})=\sum_{P}\sum_{i}u_{i}(Px_{i})$ for
bosons \cite{Girardeau:1960p10745}.
Thus, we need to solve the single-particle Schr\"odinger equation
with potential $V(x) =  (x^2/2\ell_{\mathrm{tr}}^2) +
sE_{R}\sin^{2}(qx)$, 
%
%
where $\ell_{\mathrm{tr}}$ is the trap length for the harmonic trapping
potential.  For a uniform (non-trapped) system, the solution is known to be
expressible in in terms of Mathieu functions \cite{Slater_PRB1952,
Abramowitz}.


Once the lowest $N$ single-particle orbitals $u_i(x)$ have been calculated,
the one-dimensional density can be shown to be 
\begin{equation}  \label{eq:density-bosons-fermions}
n(y)
=  \int dx_{2}...dx_{N}\left|\psi(y,x_{2},...,x_{N})\right|^2
= \sum_{i}u_{i}^{2}(y) \ .
\end{equation}
The ``filling factor'', the number of particles divided by the number of
wells, is (for uniform systems) the integrated $n(x)$ over one period:
$\nu=\int_{y}^{y+\pi/q}{dx}n(x)$.  For non-uniform systems we assign to each
well $j$ the filling $\nu(j)$, defined as the integral of $n(x)$ within that
well.

\emph{Energy scales} ---
For deep wells (large $s$), expanding $s\sin^{2}(qx)\approx sq^{2}x^{2}$ shows
that the lower energy levels have equal spacing $\propto\sqrt{s}$.  This
implies equispaced bands, which explains the equal widths ($\propto\sqrt{s}$)
of the Mott regions.  On the other hand, for $s\rightarrow0$, zero-point
energy arises due to confinement by the neighboring particles. At filling
$\nu$ this distance is $\pi/\nu{q}$ and therefore the relevant energy is
$\nu^2E_{R}$.  This explains the positions of the Mott slivers for
$s\rightarrow0$ and the linearly increasing distance between successive Mott
regions [Fig. \ref{fig_phasediagram}(d,e)].

\emph{Perturbation theory away from the Tonks-Girardeau point} ---
Near the TG limit, i.e., for finite but \emph{large} interactions, the 1D
Lieb-Liniger gas can still be mapped onto a 1D \emph{weakly}
interacting fermionic problem. The fermionic interaction is
unfortunately not simple.   We use the form \cite{Sen_pseudopotential,
  Brand:PhysRevA72:2004, brand2005dynamic, paraan, buljanlinear}
\begin{equation}
\tilde{V}(x)=-\frac{2\hbar^{4}}{m^{2}g}\delta''(x)=-g_{f}\delta''(x).\label{eq:pseudopotential}
\end{equation}
This is to be interpreted as a perturbation to the fermionic state.  Thus the
energy shift for the bosonic ground state is $\delta
E=\left<\psi_{f}\right|\tilde{V}\left|\psi_{f}\right>$.  After some
manipulation we obtain
\begin{equation}
\delta E=-\frac{g_{f}}{4}\int dR\,\left[g_{2}^{(2,0)}\left(R,R\right)-g_{2}^{(1,1)}\left(R,R\right)\right],\label{eq:delta-E}
\end{equation}
with  $g_{2}^{(m,n)}$
denoting the $m$-th ($n$-th) derivatives with respect to the first
(second) argument of the  pair correlation function  $g_{2}(x,y)$.  For the
TG gas,  $g_{2}(x,y)=n(x)n(y)-\left|\Delta(x,y)\right|^{2}$
where $\Delta(x,y)=\sum_i u_{i}^{*}(x)u_{i}(y)$, the sum running over the
first $N$ eigenstates \cite{girardeauwright}.


\begin{figure*}
\centering
\includegraphics*[width=0.99\textwidth]{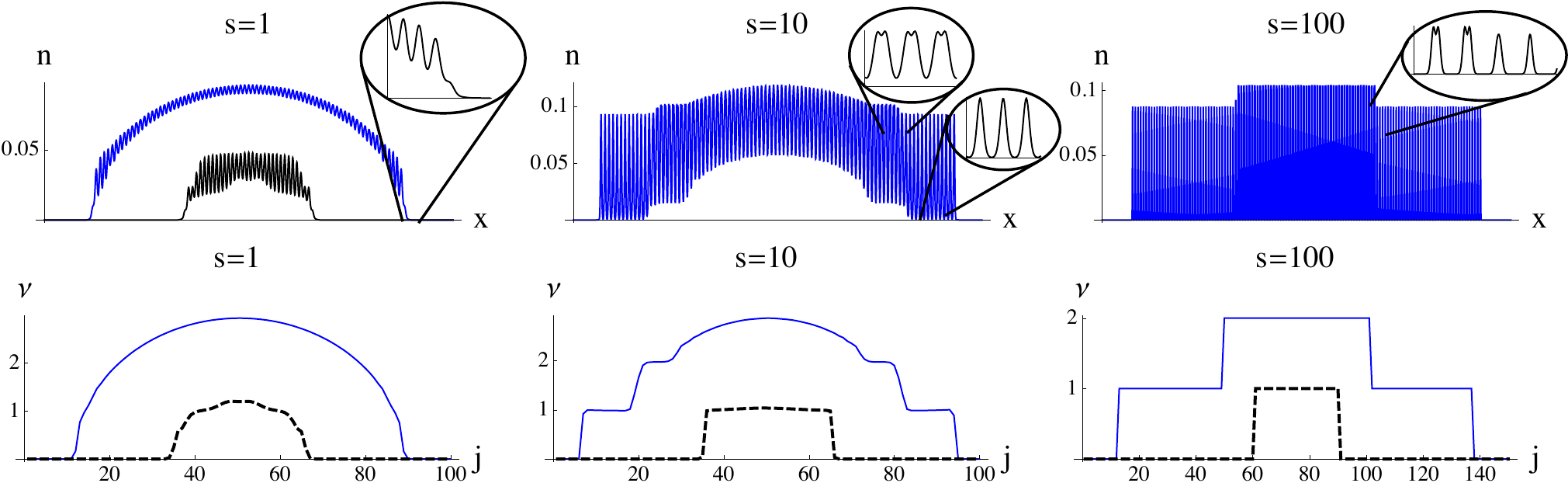}
\caption{ \label{fig_weddingcakes}
(Color online.)  A harmonic trap in addition to the modulated potential.  Top
panels: densities $n(x)$, as a function of continuous position $x$.  Bottom
panels: fillings $\nu(j)$, as a function of discrete lattice index $j$.  Blue
curves are for 181 bosons.  Lower black curves, where present, correspond to
31 bosons.  The trap length is $30/\pi$ times the modulation period.  Insets
highlight features discussed in the text.  }
\end{figure*}

\emph{Equation of state} ---
Fig.\ 2 (left) shows the chemical potential versus filling, for several different
depths.  We call these $\mu$ versus $\nu$ curves the ``equations of
state''.  As the ground state is constructed by successively occupying 
single-particle states, the $\mu(\nu)$ curves are given by the
single-particle energy dispersion curves (energy $\epsilon$ versus momentum
$k$), with the interpretation $\nu=k/q$.  
The Mott regions are incompressible in the sense that the filling does not
change with chemical potential, hence there are segments of these curves
which are vertical jumps at integer values of $\nu$.
The topography of the phase diagram --- thin Mott regions for small $s$, thick
Mott regions for large $s$ --- is visible in the structure of these $\mu(\nu)$
curves.  


We also show the perturbative correction to the equation of state curves for
$\gamma$ large but finite.  Specifically, we display $\gamma=11$,
corresponding to some of the experimental data in
Ref.~\cite{haller466pinning}.  For small and moderate lattice strengths
($s\lesssim 10$), the corrections are minute. For $s\gg10$, perturbative
results are difficult to interpret and thus not displayed.

\emph{Excitation gap in Mott phase} ---
Fig.\ 2 (right top) shows the first few energy gaps as a function of the trap
depth when the system is in the Mott phase. The gaps correspond to the vertical
jumps in the $\mu(\nu)$ curves (Fig.~\ref{fig_additional} left) or,
equivalently, to the width of the Mott regions in Fig.~\ref{fig_phasediagram}.
We find $E_g/E_R= \left| b_{\nu}(-\frac{s}{4})-a_{\nu}(-\frac{s}{4})\right|$,
with $a_{\nu}$ and $b_{\nu}$ the characteristic values of Mathieu
functions \cite{Abramowitz}.  For small $s$, the $\nu$-th gap grows as
$s^{\nu}$, e.g., as ${s/2}$ and as $s^2/32$.  For $\nu=1$ this is in agreement
with sine-Gordon results \cite{haller466pinning,kehrein_PRL99,
Buchler:2003p10714}. For $s\gg{1}$, all gaps cross over to $\sim{s}^{1/2}$, in
agreement with the energy-scale arguments given previously and consistently
with  Fig.~\ref{fig_phasediagram}(e).

The $\nu=1$ gap has been experimentally measured in \cite{haller466pinning}
through modulation spectroscopy.  In Fig.~\ref{fig_additional} (right,bottom)
we show the experimental data values (for $\gamma=11$) together with our exact
calculation for $\gamma=\infty$.  The perturbative result for $\gamma=11$ is
indistinguishable from the TG line.  Our results for trapped systems, below,
show that a uniform-system gap cannot be expected to coincide with
measurements on a trapped system with significant inhomogeneity.

\emph{An overall harmonic trap} --- 
In Fig.\ \ref{fig_weddingcakes} we show the exact density profiles of a
harmonically trapped Tonks-Girardeau gas in an optical lattice, for
various lattice depths.  
%
%
We show both the densities $n(x)$ as
a function of space and the fillings $\nu(j)$ as a function of well index.
The fillings $\nu$ are obtained by averaging $n(x)$ over each well.

The fillings (bottom row) show a wedding-cake structure familiar from the
literature on Bose-Hubbard physics in traps.  It is noteworthy that we have
obtained this from averaging densities in a continuum model, and not from a
tight-binding model like the Bose-Hubbard.  The $\nu(j)$ curves can be
understood from ``local density approximation'' arguments based on the phase
diagram presented in Fig.\ \ref{fig_phasediagram}.  Going from deeper to
shallower modulations, the Mott regions grow thinner
[Fig.\ \ref{fig_phasediagram}(d,e)], and correspondingly there are smaller plateau
regions in the density profiles.  This resembles the effect of going from
stronger to weaker interactions in the Bose-Hubbard model.

The continuum density profiles, $n(x)$, are less familiar from other contexts.
The density oscillates at the lengthscale of the well size, so we have
highlighted some features in insets.  For `fillings' larger than one, $n(x)$
within each well typically has multiple peaks.  The height of the peaks in the
$\nu=2$ plateau region is larger than, but not twice, that in the $\nu=1$
region.  Only the filling $\nu(j)$, i.e., the integrated density, has integer
values at the plateaus.  Another feature is that the densities do not
completely vanish between the wells, especially at small $s$, in neither
superfluid nor Mott region.

\begin{figure}
\centering
\includegraphics*[width=0.99\columnwidth]{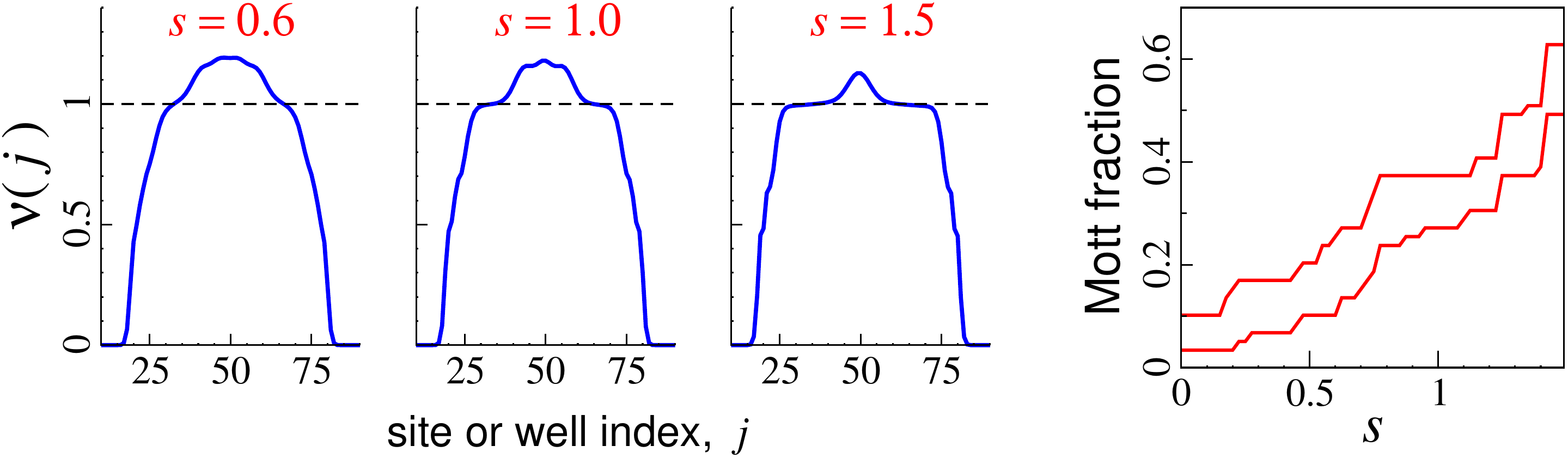}
\caption{ \label{fig_profiles_expt}
Left: Density profiles in a trap, with parameters corresponding to the central
tube of the experiment of Ref.~\cite{haller466pinning}. Right: Mott fraction,
defined as the number of sites with $|\nu-1|<\epsilon$ divided by the total
number of bosons, with $\epsilon=0.02 (0.05)$ for the lower (upper) curve.
}
\end{figure}

In Fig.\ \ref{fig_profiles_expt}, we show profiles tailored to the
experimental situation of Ref.~\cite{haller466pinning}, where the central
tubes are reported to have about 60 bosons with maximal filling around
$\nu\approx1.2$.  In a non-trapped (uniform) system at integer $\nu$, an
arbitrarily weak modulation potential pins the whole system in a Mott state.
Our results in Fig.\ \ref{fig_profiles_expt} show that the situation is more
complicated in a harmonic trap; $\nu$ is now strongly position-dependent.  In
particular, in the $s\lesssim1$ regime, the part of the cloud which is
`pinned' at $\nu=1$ is a small fraction of the total.  Therefore,
identification of modulation spectroscopy measurements~\cite{haller466pinning}
in this regime with a ``Mott gap'' is nontrivial.  Further investigation
incorporating inhomogeneity considerations is therefore called for.  While
Fig.~\ref{fig_profiles_expt} shows $\gamma=\infty$ results, the perturbative
calculations in Fig.~\ref{fig_additional} indicate that corrections to the
profiles will be tiny for interactions as low as $\gamma\sim10$.

\emph{Discussion; Open questions} ---
We have studied strongly interacting bosons at and near the hardcore limit for
periodic potentials of arbitrary depth.  We have presented the phase diagram
of the system in the $\mu$-$s$ plane, and quantitatively characterized various
related aspects.  By providing an exact and detailed map of one plane of the
$\mu$-$\gamma$-$s$ space [Fig.\ \ref{fig_phasediagram}(h)], we give a good
first idea of the topography of this 3D phase diagram, and pave the way for
future explorations of the complete 3D phase space.

Not surprisingly, the Mott regions are less dominant for weaker lattice
strengths.  This has clear consequences for trapped systems --- the fraction
of the system locked at integer values is small at small $s$.  This may
explain why the experimentally measured values for the $\nu=1$ gap fall below
the theory for the uniform system.  Heuristically, a significant part of the
system is compressible (gapless), which may be expected to reduce the measured
gap.  Since the modulation spectroscopy employed in
Ref.~\cite{haller466pinning} involves complicated temporal dynamics, this
highlights the need for studies of time dependent properties of the
non-uniform system with a weak lattice.

Other open questions raised by this work include the effects of non-infinite
$\gamma$, beyond the perturbative calculations presented here.  For $s\gg10$,
we find that the perturbative corrections at $\gamma\approx10$ lead to the
fermionic dispersion relation becoming non-monotonic, so that the Bose-Fermi
mapping would involve careful (re-)interpretation.  At small $s$, the
perturbative calculation could in principle give estimates for the $s_{\rm
crit}$ value where the Mott slivers end.  (For $\gamma^{-1}=0$ the Mott
slivers extend to $s=0$.)  In practice, we have found that the perturbative
corrections to the gap are so small that extensive high-precision computations
would be required for such a calculation.  A Monte Carlo or DMRG study, 
analogous to the 3D study of \cite{PilatiTroyer_arxiv1108}, would be
useful in this regard.

\emph{Acknowledgments} --- We thank E.~Haller and H.-C.~N\"agerl
for providing us with experimental data.

\end{document}